\documentclass[12pt,a4paper]{article}

\usepackage{natbib}
\usepackage[colorlinks,citecolor=blue,urlcolor=blue,filecolor=blue,backref=page]{hyperref}
\usepackage{graphicx,psfrag,epsf}
\usepackage{doi}
\usepackage{subfigure}
\usepackage{xcolor}
\usepackage{algorithm}
\usepackage{adjustbox}
\usepackage[noend]{algpseudocode}
\usepackage{bbm, dsfont}
\usepackage{amsfonts}
\usepackage{siunitx}
\usepackage{amsmath}
\usepackage{bbm}
\usepackage{amsfonts}
\usepackage[flushleft]{threeparttable}
\usepackage{caption}
\newcommand{\be}{\begin{equation}}
	\newcommand{\ee}{\end{equation}}
\DeclareMathOperator*{\argmax}{\arg\!\max}

\usepackage[paperheight=11in,paperwidth=10in]{geometry}

\begin{document}

	\begin{center}
		\Large \bf Estimating Prevalence of Post-war Health Disorders Using Capture-recapture Data
		
	\end{center}
	\begin{center}
		\textbf{Prajamitra Bhuyan$^{\dag}$ Kiranmoy Chatterjee$^{\ast}$    }\\
		$^{\dag}$School of Mathematical Sciences, Queen Mary University of London\\
		$^{\ast}$Department of Statistics, Bidhannagar College, Kolkata
		
	\end{center}
	
	\begin{abstract}
Effective surveillance on the long-term public health impact due to war and terrorist attacks remain limited. Such health issues are commonly under-reported, specifically for a large group of individuals. For this purpose, efficient estimation of the size of the population under the risk of physical and mental health hazards is of utmost necessity. In this context, multiple system estimation is a potential strategy that has recently been applied to quantify under-reported events allowing heterogeneity among the individuals and dependence between the sources of information. To model such complex phenomena, a novel trivariate Bernoulli model is developed, and an estimation methodology using Monte Carlo based EM algorithm is proposed. Simulation results show superiority of the performance of the proposed method over existing competitors and robustness under model mis-specifications. The method is applied to analyze real case studies on the Gulf War and 9/11 Terrorist Attack at World Trade Center, US. The results provide interesting insights that can assist in effective decision making and policy formulation for monitoring the health status of post-war survivors.	
		
		\paragraph{}	
	\end{abstract}
	
	\section{Introduction}\label{Intro}
Terrorism, as a means of intentional violence for political gains by killing and spreading fear over innocent people, becomes a serious issue for mankind specially from the last century almost all over the world. 
The 1983 Beirut barracks bombings, the 9/11 Twin Towers attacks comprising a series of airline hijackings and four coordinated suicide terrorist attacks on 2001 in the US, the 2002 Bali bombings and the 2008 Mumbai Terrorist attacks on 26/11 are some of the horrifying events to mention, but a few in this context. On an average, 24,000 people succumbed to terrorist attacks worldwide each year over the last decade \citep{Ritchie22}. On the other hand, war has long lasting history alongwith the civilization of mankind. Similar to terrorism, war involves acts of extreme violence motivated by ideological or political doctrines. However, war tends to be more widespread, and the destruction is likely to be more devastating as it is waged by states with armies equipped with lethal weapons of mass destruction. In recent times, civilian populations are more vulnerable and suffer higher casualties than professional soldiers due to the rise in civil wars and new methods of warfare \citep{COE}. Some studies suggest that 200 million people died directly or indirectly as a result of several conflicts during the twentieth century, more than half of them were civilians \citep{Victor08}. But in contemporary conflicts as many as 90\% of casualties are civilians, and the majority of them are women and children \citep{COE}. Apart from the loss of human life, a huge number of people suffer from psychological trauma, deteriorating physical and economic infrastructure, forced migration, injury, disease, lack of essential amenities like food, water or energy supplies \citep{Death20}. Experts are concerned about the spread of COVID-19 and mental health consequences for millions of people due to the Russia-Ukraine conflict \citep{IFRC22}. Apart from civilians, a significant number of soldiers suffer from post-traumatic stress disorder on returning from war zones \citep{Victor08}. In the \citet{WHA}, the WHO reported that $10\%$ of the people, who experience traumatic events due to armed conflicts, will have serious mental health problems, and another 10\% will develop behaviour that will hinder their ability to function effectively.


\subsection{Undercoverage of Humanitarian Crisis due to War and Terrorism}
The amount of losses, in terms of deaths, and physical and mental injuries caused by terrorist activity or war, are closely monitored by federal agencies. This helps to identify the actual victims who encountered such losses and formulate policies for the welfare of the survivors and the family of the deceased. Moreover, knowledge about these incidents makes the general public aware of the severity of such inhuman activities. Recent reports indicate that terrorism appears to be more savagely violent and geographically focused on the Middle East, Africa, or South Asia, as $95\%$ of deaths in 2017 occurred in those areas \citep{Ritchie22}. However, death registration systems in most of these countries are not satisfactory, which might result in a considerable amount of under-reporting of deaths and injuries in several terrorist activities and wars \citep{Kenny21}. Even in the case of a developed country like USA, the vast and varied consequences of wars and counterterror operations since the deadliest 9/11 incident were not properly accounted for \citep{BrownUniv21}. The UN Human Rights Council reported nearly 350,209 identified deaths since 2014 during the long-lasted Syria conflict from March 2011 to March 2021. Later, some leading UK-based monitoring groups indicates a notable amount of undercount in death counted by UN Human Rights Council \citep{sohr21}.  


\subsection{Motivating Case Studies}\label{motivating}
As discussed earlier, the Middle East is one of the most affected regions due to war in the last couple of decades. The Iran-Iraq war (1980-1988), the Iraqi invasion of Kuwait resulting in the Gulf War (1991), and then the US lead attack on Iraq in 2003 had a profound impact on the health of civilians and the army personnel. Around $79\%$ of the US military troop reported at least one chronic medical condition \citep{Mental06}. We consider a study on the nationwide epidemiological investigation of Amyotrophic Lateral Sclerosis (ALS) among the veteran participants in the Gulf War \citep{Gulf}. ALS is a rare neurological disease that primarily affects the nerve cells responsible for controlling voluntary muscle movements like chewing, walking, and talking. The disease is progressive, and currently, there is no cure for ALS. Due to the rarity of ALS, under-ascertainment could have a substantive effect on the rates and associated risk estimate \citep{Nelson18}. The possibility of under-ascertainment is particularly high among the non-deployed personnel who were less likely to participate in the study. Hence, the estimation of total ALS cases due to the 1991 Gulf War is important to assess the spread of the disease among military personnel. 

Another study of our present interest is associated with one of the deadliest terrorist attacks ever carried out on US soil, when four aeroplanes were hijacked on September 11, 2001. Two of the planes were flown into the twin towers of the World Trade Center (WTC) in New York City, a third plane hit the Pentagon in Arlington, Virginia, just outside Washington, D.C., and the fourth plane crashed in a field in Shanksville, Pennsylvania. In addition to the loss of life, the survivors were exposed to hazardous dust and debris from the collapsed buildings and many people suffered psychological trauma. The exact number of individuals present in the twin towers of WTC at 8:46 AM on the day of the attack is unknown. To study the long and short-term physical and mental health effects of the survivors, eligible individuals were identified through various sources and enlisted in the World Trade Center Health Registry \citep{WTCHR}. A few attempts have been made to estimate the size of the WTC population, which can provide an idea about the actual count at risk of trauma as a result of the direct exposure to that horrendous attack \citep{Cauchon07, Murphy07}.


This article considers the problem of estimating the number of victims suffering from a war or terrorist activity. In this regard, we consider two case studies on (i) ALS disease among military personnel during the 1991 Gulf War and (ii) the population in WTC Twin Towers at the time of the terrorist attacks on September 11, 2001. Very limited works have been done in the literature in this context of accounting casualties in a war or terrorist attack \citep{Coffman05,Murphy09}. The multiple system estimation (MSE) approach has a good potential to be used for estimating the size of the population under the risk of physical and mental health hazards caused due to wars or terrorist attacks \citep{Ball03,Zwierzchowski10}. In this approach, information on the bonafide victims is collected from two or more data sources to increase coverage. The group of victims remained fixed and unaltered, which ensures the basic assumption of the \textit{closed population} for MSE holds. By matching the identities from different available sources, a final estimate of the true count of casualties can be obtained. However, various modeling and estimation challenging issues are in MSE. We have discussed those in Section \ref{TRS}.

\subsection{Multiple System Estimation: Issues in Modeling and Estimation}\label{TRS}

In an MSE setup with three samples, various forms of overlap counts among the lists are summarized in the form of an incomplete $2^3$ contingency table. This data structure is typically known as the triple record system (TRS) in which, the total number of individuals with a particular capture status, say ($i,j,k$), is denoted by $x_{ijk}$, where $i$, $j$, and $k$ take value 1 for capture and 0 otherwise (see Table \ref{Tab:1}). For example, $x_{011}$ refers to the number of individuals that enlisted in the second and the third lists but are absent in the first list. Similarly, $x_{000}$ represents the number of individuals that are not included in any of the three lists and remain unknown. Therefore, the problem of estimating the population size $N=\sum_{i,j,k}x_{ijk}$ is equivalent to the estimation of unknown $x_{000}$ which refers to the number of eligible individuals not included in any of the three lists. Interested readers are referred to \cite{Darroch93,Chatterjee20a} for a detailed discussion on TRS. In each of the aforementioned case studies, the information available from different sources is summarized in the form of TRS data structure and these are presented in Table A.1 of the Supplementary Material.

\begin{table}[h]
	\begin{center}
		\caption{Data structure associated with a triple record system.}
		\begin{tabular}{lcccccccc}
			\hline
			&\multicolumn{8}{c}{List 3} \\
			\cline{2-9}
			&\multicolumn{3}{c}{In}&&&\multicolumn{3}{c}{Out} \\			
			\cline{2-9}
			&\multicolumn{3}{c}{List 2}&&&\multicolumn{3}{c}{List 2} \\	
			List 1 & In & Out & Total &&& In & Out & Total\\
			\hline \hline
			In & $x_{111}$ & $x_{101}$ & $x_{1\cdot1}$ &&& $x_{110}$ & $x_{100}$ & $x_{1\cdot0}$\\
			Out& $x_{011}$ & $x_{001}$ & $x_{0\cdot1}$ &&& $x_{010}$ & \textbf{$x_{000}$} & $x_{0\cdot0}$\\ 
			\hline
			Total& $x_{\cdot11}$ & $x_{\cdot01}$ & $x_{\cdot\cdot1}$ &&& $x_{\cdot10}$ & $x_{\cdot00}$ & $x_{\cdot\cdot0}$\\
			\hline
		\end{tabular}
		\label{Tab:1}
	\end{center}
\end{table}

The inclusion status of an event or individual in any list may have a direct causal effect on its inclusion in other lists. This phenomenon is known as \textit{list-dependence} caused by behavioral response variations in the context of MSE. Several modeling approaches for \textit{list-dependence} have been discussed in the literature \citep{Darroch93,Chao01b,Chatterjee20b}. However, modeling a capture-recapture type data becomes more challenging if there is a lack of homogeneity in capture probabilities across the individuals. This phenomenon is known as \textit{individual heterogeneity} which induces positive dependence among the lists \citep{Chandrasekar49}. In the case of disease surveillance, the interaction between lists is commonly observed over the past studies \citep{WorkingGroup95b}. In both the case studies under consideration on war and terrorism, the respective TRS datasets are potentially affected by the \textit{individual heterogeneity} as well as \textit{list-dependence} due to different demographic and ethnic backgrounds of the victims \citep{Murphy09,Gulf}. To account for the \textit{list-dependence}, log-linear models are popular in applications \citep{Fienberg72b, Cormack89}. However, \citet{Coumans17} pointed out that the parameters associated with the log-linear models are not well interpretable to the practitioners. Alternatively, the ecological model $M_{tb}$ accounts for the list dependence in terms of the behavioral response variation between the available sources that is occurred only when the inclusion status of an individual in one list has a direct causal effect on his/her inclusion status in other lists \citep{Otis78}. \citet{Chatterjee20b} discussed some other limitations of this model. In such cases, the quasi-symmetric or partial quasi-symmetric Rasch models can be used to estimate the population size \citep{Rasch61}. The most general scenario of the capture-recapture experiment encompasses both the \textit{individual heterogeneity} and \textit{list-dependence}. In this context, \citet{Chao98} proposed a non-parametric estimate of $N$ based on the sample coverage approach relaxing both the \textit{independence} as well as \textit{homogeneity} assumptions. However, the performance of this method is not satisfactory and it may provide infeasible estimates. See Section \ref{Simulation} and Section B.4 of the Supplementary Material for details. 


In this article, we propose an efficient estimation strategy of population size based on a generic model that encompasses the \textit{list-dependence} and \textit{individual heterogeneity}. The model parameters are easily interpretable and provide interesting insights into the capture-recapture setting for the human population. The details of the proposed model and its special cases are provided in Section \ref{general}. In section \ref{Inference}, estimation methodology for the population size $N$ and associated model parameters are discussed. We develop a Monte Carlo-based Expectation-Maximization algorithm for the estimation of the model parameters. In Section \ref{Simulation}, an extensive simulation study has been carried out to compare the performance of our proposed method with that of the existing competitors. Sensitivity analyses concerning different choices of underlying distributions for heterogeneous capture probabilities in the three lists are also discussed under different choices for underlying model specifications. Analyses of real datasets on ALS surveillance due to the Gulf War and Twin Towers population are presented based on our model as well as the relevant competitors in Section \ref{Casestudy}. Finally, we discuss the finding of our model towards potential applications and conclude with some discussion on future research in Section \ref{Discussion}.

\section{Modeling TRS Data}\label{general}
In a capture-recapture setup, some individuals behave independently over the three different capture attempts and behavioural dependence exists for the rest of the population. In this context, \citet{Chatterjee20b} proposed a Trivariate Bernoulli model (TBM) with nice physical interpretations to account for such inherent dependence among the lists. Denote $(Z_h^{(1)}, Z_h^{(2)},Z_h^{(3)})$ and $(X_{1h}^{*},X_{2h}^{*},X_{3h}^{*})$ as the actual and the latent inclusion statuses of the $h$th individual, respectively, in three lists $L_1$, $L_2$, and $L_3$, for $h=1,2,\ldots,N$. The capture statuses $Z_h^{(l)}$ and $X_{lh}^{*}$ take value 1 or 0, denoting the presence or absence of the $h$th individual in the $l$th list, for $l=1,2,3$. Under this setup, the pairwise positive dependence between lists ($L_1$ and $L_2$), ($L_2$ and $L_3$) and ($L_1$ and $L_3$) are modeled as $X_{2h}^{*}=X_{1h}^{*}$, $X_{3h}^{*}=X_{2h}^{*}$, and $X_{3h}^{*}=X_{1h}^{*}$ for $\alpha_{1}$, $\alpha_{2}$, and $\alpha_{3}$ proportion of individuals, respectively. As example, for each of all the $\alpha_{1}$ proportion of individuals in the given population, inclusion status in List 1 (i.e. presence or absence) positively influences his/her inclusion status in List 2. Similarly, second-order dependency among the three lists $L_1$, $L_2$ and $L_3$ is accounted by considering $X_{3h}^{*}=X_{2h}^{*}=X_{1h}^{*}$ for $\alpha_{4}$ proportion of individuals which represents joint positive interaction among the three lists. Therefore, the remaining $(1-\alpha_0)$ proportion of individuals, where $\alpha_0=\sum_{l=1}^{4}\alpha_{s}$, behave independently over the three lists, and we can formally write the model to account the interdependencies among the three lists as:
\begin{eqnarray}
	(Z_h^{(1)},Z_h^{(2)}, Z_h^{(3)}) = \begin{cases} 
		(X_{1h}^{*},X_{1h}^{*},X_{3h}^{*})  & \mbox{ with prob. } \alpha_{1},\\
		(X_{1h}^{*},X_{2h}^{*},X_{2h}^{*})  & \mbox{ with prob. } \alpha_{2},\\
		(X_{1h}^{*},X_{2h}^{*},X_{1h}^{*})  & \mbox{ with prob. } \alpha_{3},\\
		(X_{1h}^{*},X_{1h}^{*},X_{1h}^{*})  & \mbox{ with prob. } \alpha_{4},\\
		(X_{1h}^{*},X_{2h}^{*},X_{3h}^{*}) & \mbox{ with prob. } 1-\alpha_0,
	\end{cases}\label{prob-model}
\end{eqnarray}  
where $X_{1h}^{*}$'s, $X_{2h}^{*}$'s and $X_{3h}^{*}$'s are independently distributed Bernoulli random variables with parameters $\mathcal{P}_{1}$, $\mathcal{P}_{2}$ and $\mathcal{P}_{3}$, respectively, for all $h=1,\ldots,N$. Note that $\mathcal{P}_{l}$ refers to the capture probability of a causally independent individual in the $l$th list. To extend this model in a more general setup to account for the \textit{individual heterogeneity} in addition to the \textit{list-dependence}, as discussed in section \ref{TRS}, we consider $X_{lh}^{*}$'s to follow independent Bernoulli distributions with parameter $\mathcal{P}_{lh}$,
where $\mathcal{P}_{lh}$'s are independent and identically distributed realizations of from a distribution with support $(0,1)$, for each $l=1,2,3$. 
We refer to this generic model as the Trivariate Heterogeneous Bernoulli model (THBM) which incorporates \textit{individual heterogeneity} and \textit{behavioral dependence} arising from different sources.

\subsection{Special Cases}
\citet{Chao01a} discussed the fact that the effect of the behavioral response variation is possibly confounded with the effect of heterogeneous catchability. In such cases, the lists may become positively dependent whenever the capture probabilities are heterogeneous across individuals in all the lists even if the individual capture statuses are independent \citep{Chandrasekar49}. To model such scenarios, one can consider a parsimonious model similar to that of a generalized Rasch model \citep{Darroch93,Chao98} (see Section B.3 of the Supplementary Material) with $\alpha_s=0$, for $s=1, \ldots,4$ in THBM. One can further assume $\mathcal{P}_{l}$'s are identically distributed where the catchability of the individuals is the same irrespective of the lists. The THBM model reduces to TBM when $\mathcal{P}_{l}$'s are assumed as fixed effects. This model is useful for TRS when the individuals are equally catchable in each list. It is also noted that TBM and the $M_t$ model \citep{Otis78} are equivalent in the absence of \textit{list-dependence} (i.e. $\alpha_s=0$, for $s=1,\ldots,4$). 

\section{Estimation Methodology}\label{Inference}
Traditionally in the context of MSE, population size $N$ is estimated based on the likelihood theory, where the vector of observed cell counts $$\boldsymbol{x}=\left\{x_{ijk}:x_{ijk}=\sum_{h=1}^{n}\mathbb{I}\left[Z_h^{(1)}=i, Z_h^{(2)}=j, Z_h^{(3)}=k\right]; i,j,k=0,1;i=j=k\neq0\right\}$$ (as presented in Table \ref{Tab:1}) follow a multinomial distribution with index parameter $N$ and the associated cell probabilities $\boldsymbol{p}=\{p_{ijk}: i,j,k=0,1;i=j=k\neq0\}$ \citep{Sanathanan72}, where $\mathbb{I}\left[\cdot\right]$ denotes indicator function. Therefore, the likelihood function is given by
\begin{eqnarray}\label{Mult_lik}
	L(N,\boldsymbol{p}|\boldsymbol{x})& = & \frac{N!}{\prod_{i,j,k=0,1;i=j=k\neq0}^{}x_{ijk}!(N-x_{0})!}\prod_{i,j,k=0,1}^{}p_{ijk}^{x_{ijk}},\nonumber
\end{eqnarray}
where $x_{000}=N-x_0$. We can rewrite the likelihood function based on the TBM using its relationships with the cell probabilities $p_{ijk}$, as
\begin{eqnarray}\label{L_Model_I}
	L(N,\boldsymbol{\alpha},\boldsymbol{\mathcal{P}}|\boldsymbol{x})&\propto & \frac{N!}{(N-x_{0})!} [(1-\alpha_0)\mathcal{P}_1 \mathcal{P}_2 \mathcal{P}_3 + \alpha_1 \mathcal{P}_1 \mathcal{P}_3 + \alpha_2 \mathcal{P}_1 \mathcal{P}_2 + \alpha_3 \mathcal{P}_1 \mathcal{P}_2 + \alpha_4 \mathcal{P}_1]^{x_{111}}\nonumber\\
	&&\times
	[(1-\alpha_0) \mathcal{P}_1 \mathcal{P}_2 (1-\mathcal{P}_3) + \alpha_1 \mathcal{P}_1 (1-\mathcal{P}_3)]^{x_{110}}\nonumber\\
	&&\times
	[(1-\alpha_0)(1-\mathcal{P}_1)\mathcal{P}_2 \mathcal{P}_3 + \alpha_2 (1-\mathcal{P}_1) \mathcal{P}_2]^{x_{011}}\nonumber\\
	&&\times
	[(1-\alpha_0)\mathcal{P}_1(1-\mathcal{P}_2)(1-\mathcal{P}_3) + \alpha_2 \mathcal{P}_1 (1-\mathcal{P}_2)]^{x_{100}}\nonumber\\
	&&\times
	[(1-\alpha_0)\mathcal{P}_1(1-\mathcal{P}_2)\mathcal{P}_3 + \alpha_3 \mathcal{P}_1 (1-\mathcal{P}_2)]^{x_{101}}\nonumber\\  
	&&\times
	[(1-\alpha_0)(1-\mathcal{P}_1)\mathcal{P}_2(1-\mathcal{P}_3) + \alpha_3 (1-\mathcal{P}_1) \mathcal{P}_2]^{x_{010}}\nonumber\\
	&&\times
	[(1-\alpha_0)(1-\mathcal{P}_1)(1-\mathcal{P}_2)\mathcal{P}_3 + \alpha_1 (1-\mathcal{P}_1) \mathcal{P}_3]^{x_{001}}\nonumber\\
	&&\times
	[(1-\alpha_0) (1-\mathcal{P}_1) (1-\mathcal{P}_2) (1-\mathcal{P}_3) + \alpha_1 (1-\mathcal{P}_1) (1-\mathcal{P}_3)\nonumber\\ 
	&& +  \alpha_2 (1-\mathcal{P}_1) (1-\mathcal{P}_2) + \alpha_3 (1-\mathcal{P}_1) (1-\mathcal{P}_2)+ \alpha_4 (1-\mathcal{P}_1)]^{N-x_{0}},
\end{eqnarray}
where $x_{000}=N-x_{0}$, $p_{000}=1-\sum_{i,j,k=0,1;i=j=k\neq0}^{}p_{ijk}$, $\boldsymbol{\alpha}=(\alpha_1, \alpha_2, \alpha_3, \alpha_4)$, and $\boldsymbol{\mathcal{P}}=(\mathcal{P}_1, \mathcal{P}_2, \mathcal{P}_3)$. 
In addition to the first and second order \textit{list-dependence} involved in TBM, the proposed THBM accounts for the \textit{individual heterogeneity} in terms of the variations in capture probabilities considering $\mathcal{P}_l$ as a random effect for $l=1,2,3$.  Now, integrating the likelihood function, given in (\ref{L_Model_I}), with respect to the distribution of  $\mathcal{P}_{l}$, we obtain the marginal likelihood function of $\boldsymbol{\theta}$ as
\begin{eqnarray}\label{ultimate_likelihood}
	L(\boldsymbol{\theta}|\boldsymbol{x})& = & \int_{\mathbb{R}^3}^{} L(N,\boldsymbol{\alpha},\boldsymbol{\mathcal{P}}|\boldsymbol{x})\times\left\{\prod_{l=1}^{3} g_{\mathcal{P}_l}(\mathcal{P}_{l}|\delta_{l})\right\}d\mathcal{P}_{1}d\mathcal{P}_{2}d\mathcal{P}_{3},
\end{eqnarray}
where $g_{\mathcal{P}_l}(\cdot|\delta_l)$ is the density of $\mathcal{P}_l$ with parameter $\delta_l$ for $l=1,2,3$, and $\boldsymbol{\theta}=(N,\boldsymbol{\alpha},\boldsymbol{\delta})$.
The estimation of $N$ and other associated parameters in THBM is computationally challenging in a conventional frequentist setup due to the involvement of intractable numerical integrals \citep{Coull99}. One can consider approximation methods for numerical integration like Monte Carlo or Laplace method. 

\subsection{Identifiability Issues}\label{Ident}
In MSE, it is quite common to encounter identifiability issues with data available from a small number of dependent sources. To deal with such issues, some restrictions on parameters are considered in the existing models like $M_{tb}$ and log-linear models as well as in the sample coverage approach proposed by \citet{Chao98}. See Section B of the Supplementary Material for details. Similarly for TBM, \citet{Chatterjee20b} considered two submodels TBM-1 and TBM-2 keeping $\alpha_3$ and $\alpha_4$ fixed as $0$, respectively. However, this process of model contraction may lead to a model representing unrealistic scenarios \citep{Gustafson05}. Unlike the situation that is modeled by TBM-1, available sources related to the surveillance of humanitarian crisis are not time ordered \citep{Chan2021}, and the capture attempts are generally interdependent between themselves, including the dependence between the first and third lists \citep{Gold15}. Similarly, TBM-2 is not suitable for the scenarios when the second-order interaction among the lists is present. The proposed THBM is intended to model more complex scenarios compared to the existing models and it also suffers from similar identifiability issues with respect to TRS data. It is important to note that a non-identifiable model may bring information about the parameters of interest, and identifiable models do not necessarily lead to better decision-making than non-identifiable ones \citep{Gustafson05, Wechsler13}. Nevertheless, it is in the best interest of the practitioners, clear guidelines should be provided that are applicable to different scenarios under consideration. In the following subsection, we propose an estimation methodology of the parameter of interest $N$ and other associated parameters of the THBM based on the MCEM algorithm which provides a generic solution to the identifiability issue.

\subsection{MCEM Algorithm}\label{MCEM}
As discussed in section \ref{Inference}, the likelihood function $L(\boldsymbol{\theta}|\boldsymbol{x})$, given by (\ref{ultimate_likelihood}), is not mathematically tractable due to involvement of an integral and the additive structure of the cell probabilities ${p_{ijk}}$ as a function of the model parameters. To simplify the likelihood function we adopt data augmentation strategy proposed by \citet{Tanner87}. First, we partition the observed cell counts $x_{ijk}$ depending on the various types of \textit{list-dependence} as described in (\ref{prob-model}), and define a vector of latent cell counts as
$${\boldsymbol{y}=\left\{y_{ijk,u}: \sum_{u=1}^{\nu}y_{ijk,u}=x_{ijk}; i,j,k=0,1; \nu=5^{\omega}3^{1-\omega}; \omega= \mathbb{I}[i=j=k] \right\}},$$
where $\mathbb{I}[\cdot]$ is an indicator function. For example, $x_{111}$ can arise from all the five types of dependence structures, whereas $x_{110}$ can be generated only based on the first and last types of dependence structures, as presented in (\ref{prob-model})
We also treat the capture probabilities $\boldsymbol{\mathcal{P}}=(\mathcal{P}_1,\mathcal{P}_2,\mathcal{P}_3)$ as unobserved data. Therefore, the likelihood function of $\boldsymbol{\theta}$, based on the complete data $(\boldsymbol{x},\boldsymbol{y},\boldsymbol{\mathcal{P}})$, is given by 
\begin{eqnarray}\label{com_likelihood}
	\mathcal{L}_c(\boldsymbol{\theta}|\boldsymbol{x},\boldsymbol{y},\boldsymbol{\mathcal{P}})& = & \left\{K(\boldsymbol{y)}\right\}^{-1}  \mathcal{L}_c(N,\boldsymbol{\alpha},\boldsymbol{\mathcal{P}}|\boldsymbol{x},\boldsymbol{y})\times\left\{\prod_{l=1}^{3} g_{\mathcal{P}_l}(\mathcal{P}_{l}|\delta_{l})\right\},
\end{eqnarray}
where
\begin{eqnarray}\label{Hcom_like}
	\mathcal{L}_c(N,\boldsymbol{\alpha},\boldsymbol{\mathcal{P}}|\boldsymbol{x},\boldsymbol{y})&\propto & \frac{ N!}{(N-x_{0})!} [(1-\alpha_0)\mathcal{P}_1 \mathcal{P}_2 \mathcal{P}_3]^{y_{111,1}}\times[\alpha_1 \mathcal{P}_1 \mathcal{P}_3]^{y_{111,2}}\times [\alpha_2 \mathcal{P}_1 \mathcal{P}_2]^{y_{111,3}}\times[\alpha_3 \mathcal{P}_1 \mathcal{P}_2]^{y_{111,4}}\nonumber\\
	&&\times[\alpha_4 \mathcal{P}_1]^{x_{111}-\sum_{i=1}^{4}y_{111,i}}\times
	[(1-\alpha_0) \mathcal{P}_1 \mathcal{P}_2 (1-\mathcal{P}_3)]^{y_{110,1}} \times [\alpha_1 \mathcal{P}_1 (1-\mathcal{P}_3)]^{x_{110}-y_{110,1}}\nonumber\\
	&&\times
	[(1-\alpha_0)(1-\mathcal{P}_1)\mathcal{P}_2 \mathcal{P}_3]^{y_{011,1}} \times [\alpha_2 (1-\mathcal{P}_1) \mathcal{P}_2]^{x_{011}-y_{011,1}}\nonumber\\
	&&\times
	[(1-\alpha_0)\mathcal{P}_1(1-\mathcal{P}_2)(1-\mathcal{P}_3)]^{y_{100,1}} \times [\alpha_2 \mathcal{P}_1 (1-\mathcal{P}_2)]^{x_{100}-y_{100,1}}\nonumber\\
	&&\times
	[(1-\alpha_0)\mathcal{P}_1(1-\mathcal{P}_2)\mathcal{P}_3]^{y_{101,1}} \times [\alpha_3 \mathcal{P}_1 (1-\mathcal{P}_2)]^{x_{101}-y_{101,1}}\nonumber\\  
	&&\times
	[(1-\alpha_0)(1-\mathcal{P}_1)\mathcal{P}_2(1-\mathcal{P}_3)]^{y_{010,1}} \times [\alpha_3 (1-\mathcal{P}_1) \mathcal{P}_2]^{x_{010}-y_{010,1}}\nonumber\\
	&&\times
	[(1-\alpha_0)(1-\mathcal{P}_1)(1-\mathcal{P}_2)\mathcal{P}_3]^{y_{001,1}} \times [\alpha_1 (1-\mathcal{P}_1) \mathcal{P}_3]^{x_{001}-y_{001,1}}\nonumber\\
	&&\times
	[(1-\alpha_0) (1-\mathcal{P}_1) (1-\mathcal{P}_2) (1-\mathcal{P}_3)]^{y_{000,1}} \times [\alpha_1 (1-\mathcal{P}_1) (1-\mathcal{P}_3)]^{y_{000,2}}\nonumber\\ 
	&& \times [\alpha_2 (1-\mathcal{P}_1) (1-\mathcal{P}_2)]^{y_{000,3}} \times [\alpha_3 (1-\mathcal{P}_1) (1-\mathcal{P}_2)]^{y_{000,4}}\nonumber\\ &&\times [\alpha_4 (1-\mathcal{P}_1)]^{N-x_{0}-\sum_{i=1}^{4}y_{000,i}},\nonumber
\end{eqnarray}
and
\begin{eqnarray}
	K(\boldsymbol{y})&=&\prod_{u=1}^{4}y_{111,u}!(x_{111}-\sum_{u=1}^{4}y_{111,u})! y_{110,1}!(x_{110}-y_{110,1})! \nonumber\\ 
	&& \times y_{011,1}!(x_{011}-y_{011,1})! y_{100,1}!(x_{100}-y_{100,1})! \nonumber\\
	&& \times y_{101,1}!(x_{101}-y_{101,1})! 
	y_{010,1}! (x_{010}-y_{010,1})! \nonumber\\
	&& \times y_{001,1}!(x_{001}-y_{001,1})!
	\prod_{u=1}^{4}y_{000,u}!(N-x_{0}-\sum_{u=1}^{4}y_{000,u})!\nonumber.
\end{eqnarray}
Interestingly, the complete data likelihood, given by (\ref{com_likelihood}), possesses a simple form as a product of the power functions of the parameters associated with the THBM. However, this likelihood function cannot be used for estimation purposes without resolving the identifiability issues. For this purpose, we employ a similar approach as suggested by \citet{Carle_1978} to obtain a likelihood function free of the nuisance parameters. We consider $\mathcal{P}_l$'s follow independent beta distribution with shape parameters $n_{l}$ and $m_{l}$, for $l=1,2,3$, and obtain their estimates as $\hat{m}_1=x_{1\cdot\cdot}$, $\hat{n}_1=(N-x_{1\cdot\cdot})$, $\hat{m}_2=y_{111,1}+y_{111,3}+y_{111,4}+y_{110,1}+x_{011}+x_{010}$, $\hat{n}_2=x_{100}+x_{101}+y_{001,1}+y_{000,1}+y_{000,3}+y_{000,4}$, $\hat{m}_3=y_{111,1}+y_{111,2}+y_{011,1}+y_{101,1}+x_{001}$ and $\hat{n}_3=x_{110}+y_{100,1}+y_{010,1}+y_{000,1}+y_{000,2}$, by matching moments based on the observed and latent cell counts. Then plugging the corresponding density function of $\mathcal{P}_l$ in (\ref{com_likelihood}), we get 
\begin{eqnarray}\label{new_com_likelihood}
	\mathcal{L}_c(\boldsymbol{\tilde\theta}|\boldsymbol{x},\boldsymbol{y},\boldsymbol{\mathcal{P}})& = & \left\{K(\boldsymbol{y)}\right\}^{-1}  \mathcal{L}_c(N,\boldsymbol{\alpha},\boldsymbol{\mathcal{P}}|\boldsymbol{x},\boldsymbol{y})\times\left\{\prod_{l=1}^{3} \frac{ {\mathcal{P}_l}^{\hat{m}_{l}-1}\left[1-\mathcal{P}_l\right]^{\hat{n}_{l}-1}}{B(\hat{m}_{l}, \hat{n}_{l})}\right\},
\end{eqnarray}
where $\boldsymbol{\tilde\theta}=(N,\boldsymbol{\alpha})$.
To implement the MCEM algorithm, one must be able to sample from the conditional distributions of $\boldsymbol{y}$ given $(\boldsymbol{x}, N,\boldsymbol{\alpha}, \boldsymbol{\mathcal{P}})$, and $\boldsymbol{\mathcal{P}}$ given $(\boldsymbol{x},\boldsymbol{y}, N,\boldsymbol{\alpha})$, denoted by $\pi(\boldsymbol{y}|\boldsymbol{x}, N,\boldsymbol{\alpha}, \boldsymbol{\mathcal{P}})$, and $\pi(\boldsymbol{\mathcal{P}}|\boldsymbol{x},\boldsymbol{y}, N,\boldsymbol{\alpha})$, respectively. Note that $\pi(\boldsymbol{\mathcal{P}}|\boldsymbol{x},\boldsymbol{y}, N,\boldsymbol{\alpha})$ is expressed as the product of Beta distributions, and $\pi(\boldsymbol{y}|\boldsymbol{x}, N,\boldsymbol{\alpha}, \boldsymbol{\mathcal{P}})$ is expressed as a product of Multinomial and Binomial distributions (see Section C.2 of the Supplementary Material for details). This approach is close in spirit to the method of \textit{EM-within-Gibbs} and stochastic \textit{EM-within-Gibbs} considered by \citet{Chatterjee16b}. Now, we employ a simple iterative algorithm to estimate the model parameters using the following steps.

\begin{itemize}
	\item[Step 1.] Set $t=0$ and initialize $(\boldsymbol{y}^{(t)}, N^{(t)}, \boldsymbol{\alpha}^{(t)})$.
	
	
	\item[Step 2.] Generate $K$ samples
	$\boldsymbol{\mathcal{P}_{i}}^{(t+1)}$ from $\pi(\boldsymbol{\mathcal{P}}|\boldsymbol{x},\boldsymbol{y}^{(t)}, N^{(t)},\boldsymbol{\alpha}^{(t)})$, and then $\boldsymbol{y}^{(t+1)}_{i}$ from\\  $\pi(\boldsymbol{y}|\boldsymbol{x},N^{(t)},\boldsymbol{\alpha}^{(t)}, \boldsymbol{\mathcal{P}}^{(t+1)})$ for $i=1,\ldots,K$.
	
	\item[Step 3.] E step: Monte Carlo approximation of (\ref{new_com_likelihood}) is computed as  
	$$\frac{1}{K}\sum_{i=1}^{K} \log\mathcal{L}_c(\boldsymbol{\tilde\theta}|\boldsymbol{x},\boldsymbol{y}^{(t+1)}_{i},\boldsymbol{\mathcal{P}}^{(t+1)}_{i})$$
	\item[Step 4.] M Step:  $\underset{\boldsymbol{\tilde{\theta}^{(t+1)}} }{\argmax} \quad \frac{1}{K}\sum_{i=1}^{K} \log\mathcal{L}_c(\boldsymbol{\tilde\theta}|\boldsymbol{x},\boldsymbol{y}^{(t+1)}_{i},\boldsymbol{\mathcal{P}}^{(t+1)}_{i})$
	\item[Step 5.] Repeat Step 2-4 until convergence of $\left\{\boldsymbol{\tilde{\theta}}^{(t)}\right\}_{t\ge 0}$.
\end{itemize}

\section{Simulation Study}\label{Simulation}
In this section, we evaluate the performance and robustness of the proposed estimator through simulations. We consider eight different compositions of populations representing varying degree of \textit{list-dependence} and \textit{individual heterogeneity} in capture probabilities and denote them as P1-P8 in Table \ref{Population}. The choices of $\boldsymbol{\alpha}$ represent different degrees of \textit{list-dependence}. The heterogeneity in capture probabilities are modeled as $b_{l}=\log\left[\frac{\mathcal{P}_{l}}{1-\mathcal{P}_{l}}\right]$, where $b_{l}$'s are normal and generalized logistic type-I variates for $l=1,2,3$. To compare the performance of the proposed estimator (THBM) with the existing competitors, we consider the following models: log-linear model (LLM) \citep{Fienberg72b}, $M_{tb}$ model \citep{Chao00,Chatterjee20b}, quasi-symmetry model (QSM) \citep{Darroch93}, partial quasi-symmetry model (PQSM) \citep{Darroch93}, non-parametric sample coverage method (SC) \citep{Chao98} and the independent model (IM) i.e. LLM without interaction effects \citep{Fienberg72b}. The details of these existing models and associated estimates are briefly presented in Section B of the Supplementary Material.

\begin{figure}[htp]
	\centering
	\includegraphics[scale=1.1]{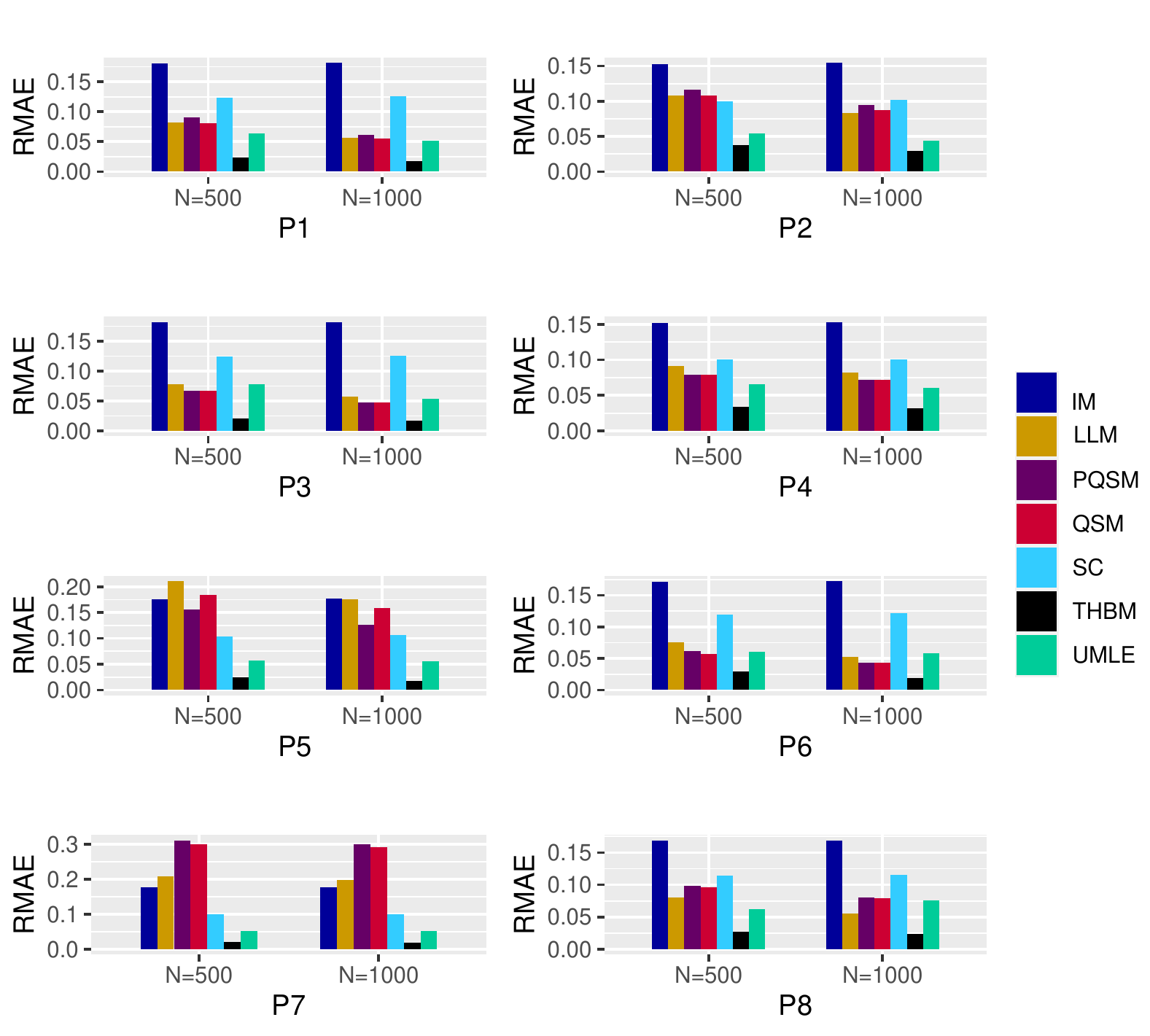}
	\caption{Comparison of RAMEs of the estimators for populations P1-P8.}
	\label{RAME}
\end{figure}

\begin{figure}[htp]
	\centering
	\includegraphics[scale=1.1]{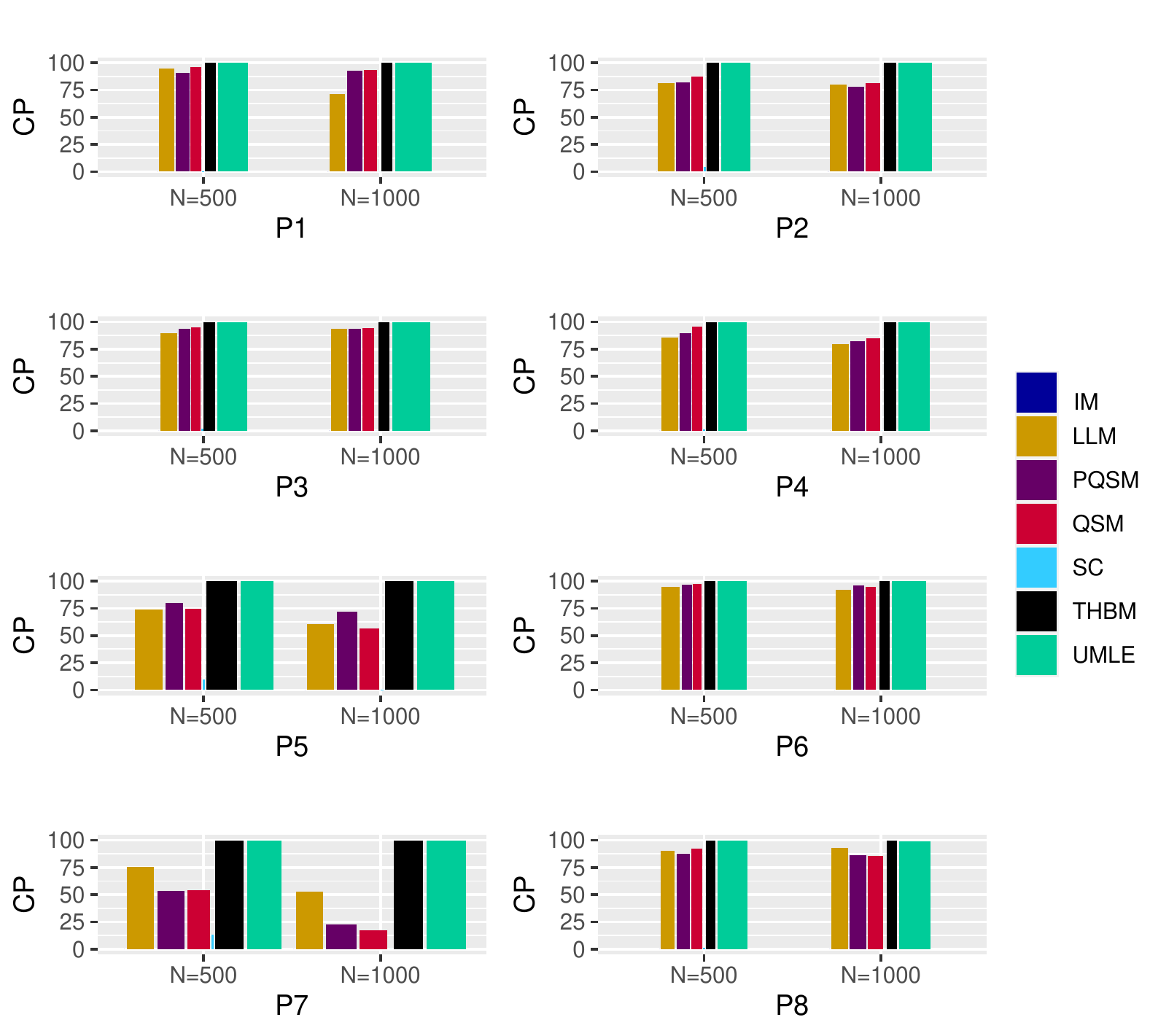}
	\caption{Comparison of CPs (in \%) and the relative LCIs of the estimators, represented by the bar height and bar width, respectively, for populations P1-P8.}
	\label{CP}
\end{figure}

\begin{table}[h]
	\begin{center}
		\caption{Composition of the simulated populations.}
		\begin{threeparttable}
			\begin{tabular}{|c|c|c|c|c|c|}
				\hline
				Populations & $(\alpha_1,\alpha_2,\alpha_3,\alpha_4)$ &  $b_1$ & $b_2$ & $b_3$\\
				\hline
				P1  &  $(0.30,0.30,0.15,0.10)$ &  $Normal(1,5)$ & $Normal(0.5,5)$ & $Normal(0,5)$  \\
				P2  &  $(0.30,0.30,0.15,0.10)$ &    $Normal(0.5,1)$ & $Normal(0.4,1)$ & $Normal(0.3,1)$  \\
				P3  &  $(0.25,0.15,0.35,0.10)$ &  $Normal(1,5)$ & $Normal(0.5,5)$ & $Normal(0,5)$  \\
				P4  &  $(0.25,0.15,0.35,0.10)$ & $Normal(0.5,1)$ & $Normal(0.4,1)$ & $Normal(0.3,1)$  \\
				P5  &  $(0.30,0.30,0.15,0.10)$ &    $GL_I(1)$ & $GL_I(1.4)$ & $GL_I(1.8)$ \\       
				P6  &  $(0.30,0.30,0.15,0.10)$ &    $GL_I(1.6)$ & $GL_I(1.2)$ & $GL_I(0.8)$ \\ 
				P7  &  $(0.25,0.15,0.35,0.10)$ &    $GL_I(1)$ & $GL_I(1.4)$ & $GL_I(1.8)$ \\       
				P8  &  $(0.25,0.15,0.35,0.10)$ &    $GL_I(1.6)$ & $GL_I(1.2)$ & $GL_I(0.8)$ \\ 
				\hline
			\end{tabular}
			\begin{tablenotes}
				\tiny
				\item $Normal(\mu, \sigma)$ denotes normal distribution with mean $\mu$ and standard deviation $\sigma$.
				\item $GL_{I}(\eta)$ denotes generalized logistic type-I distribution with parameter $\eta$.
			\end{tablenotes}
		\end{threeparttable}
		\label{Population}
	\end{center}
\end{table}
We generate 1000 datasets from the proposed model keeping the total population size $N$ fixed at 500 and 1000 for each of these eight different choices of populations P1-P8 (see Table \ref{Population}), and compute the relative Mean Absolute Error as $RMAE =\frac{1}{1000}\sum_{r=1}^{1000}|\frac{\hat{N}_r-N}{N}|$, where $\hat{N}_r$ denotes the estimate based on dataset generated at the $r$th replication, for $r=1,\ldots,1000$. The comparison of different estimators with respect to RAME is presented in Figure \ref{RAME}. In the capture-recapture setting, point estimators of population size are commonly possess positively skewed distributions and we also observed similar pattern in our study. Therefore, we obtain $95\%$ interval (C.I.) for $N$ based on the log-transformation method, proposed by \citet{Chao87}. In this method, $\log(\hat{N}-x_{0})$ is approximately treated as normal variate and that gives $95\%$ confidence interval for $N$ as 
${[x_{0}+(\hat{N}-x_{0})/C}, {x_{0}+(\hat{N}-x_{0})C ]}$,
where $\textstyle{C=\exp\left[1.96\times\left[\log(1+\hat{\sigma}_{\hat{N}}^2/(\hat{N}-x_{0})^2)\right]^{1/2}\right]}$, and $\hat{\sigma}_{\hat{N}}^2$ is the estimate of the variance of the estimator $\hat{N}$, computed based on 1000 nonparametric bootstrap samples which are drawn from implicitly assumed multinomial distribution for the mark-recapture model \citep{Buckland91}. Following \citet{Chatterjee18}, we compute the length of the $95\%$ confidence interval (LCI) as well as its coverage probability (CP) and represent both the results in Figure \ref{CP}, where height and width of each bar represent CP and LCI of the corresponding method, respectively.

The proposed model performs the best in terms of RAME for all the populations P1-P8 for both $N=500$ and $1000$. Moreover, the CPs of the proposed THBM estimator are higher than all other competitors. The CPs of $M_{tb}$ are similar to THBM but CIs are very wide due to high variability. However, the LCIs of the proposed THBM estimate are the smallest in all the cases under consideration except P5 and P7. As expected, the RAMEs of the IM are the highest among all the estimators for most of the populations. The performance of LLM, QSM and PQSM estimators are comparable with respect to both RMAE and CP. But the $M_{tb}$ performs better than LLM, QSM and PQSM. It is important to note that the CPs of SC and IM are very low as the associated CIs are very tight. The performances of each estimator are similar over all the populations with respect to CPs and LCIs except P5 and P7, where the LCIs are comparably wider. We also observe that the SC estimate suffers from boundary problems (see Section B.4 of the Supplementary Material for details) in our simulation study. One can easily configure populations to generate data such that the percentage of infeasible estimates obtained from SC method can be as high as 60\% to 80\%. Also, the QSM, PQSM, and LLM may fail to converge in some situations and produce extremely large estimates. We have not considered those cases here for a valid and fair comparison.

\subsection{Sensitivity of Model Misspecification}
\begin{figure}[htp]
	\centering
	\includegraphics[scale=1.1]{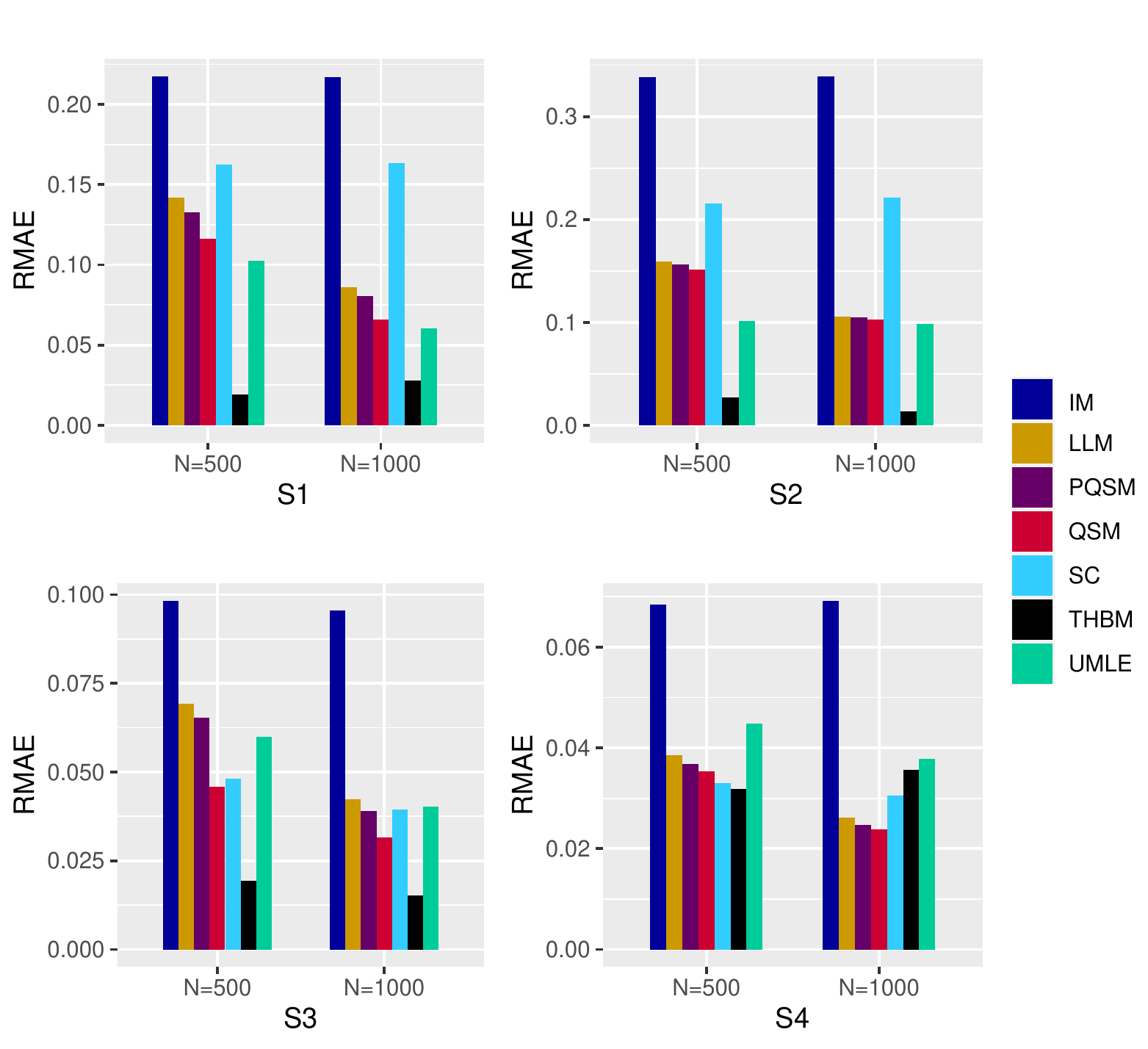}
	\caption{Comparison of the RAMEs of the proposed estimator with the existing competitors under model misspecification when data generated from configurations S1-S4.}
	\label{RAME_S}
\end{figure}

\begin{figure}[htp]
	\centering
	\includegraphics[scale=1.1]{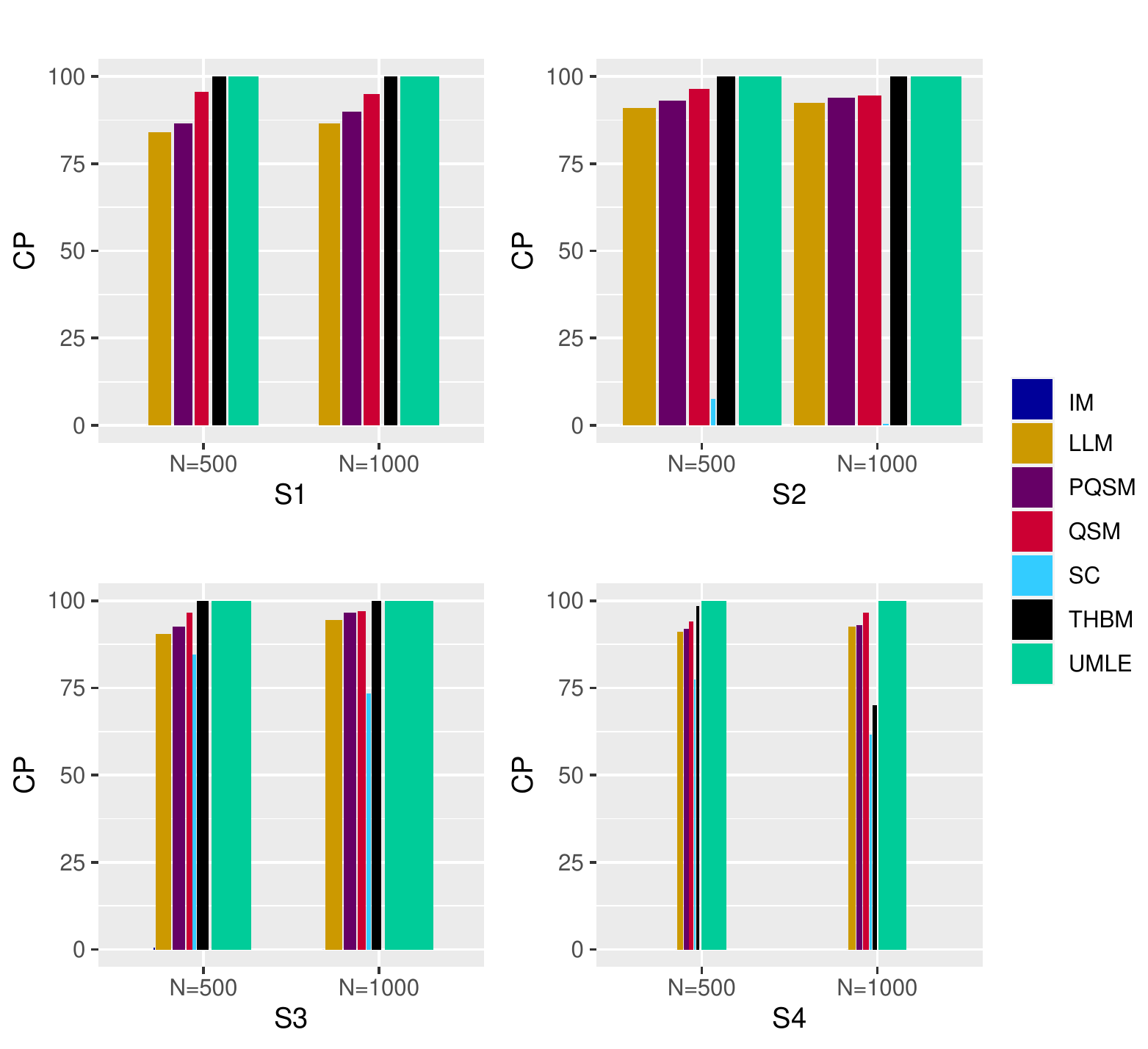}
	\caption{Comparison of the CPs (in \%) of the proposed estimator with the existing competitors under model misspecification when data generated from configurations S1-S4.}
	\label{CP_S}
\end{figure}

To study the effect of model misspecification when the data generating mechanism deviates from the fitted model, we analyse sensitivity of the estimates based on the proposed THBM in comparison with the existing estimators.
For this purpose, we generate the latent capture statuses of the individuals
$X_{1h}^{*}$, $X_{2h}^{*}$, $X_{3h}^{*}$ from Bernoulli distributions with respective probabilities
$P_{h}^{(1)}$, $P_{h}^{(2)}$, and $P_{h}^{(3)}$, where $P_{h}^{(j)}= \min\{P_{h}^{(j-1)}{\mathbb{I}[Z_{h}^{(j-1)}=0]}+1.2{\mathbb{I}[Z_{h}^{(j-1)}=1]}, 0.99\}$, for $j=2,3$, and $h=1,\ldots, N$,
and then generate $Z_{h}^{(1)}$, $Z_{h}^{(2)}$, $Z_{h}^{(3)}$ using (\ref{prob-model}) with $\boldsymbol{\alpha}=(0.1,0.1,0.1,0.1)$. This mechanism induces an additional dependence among the capture statuses similar to a first-order auto-regressive model as considered in \citet{Chao87}. To incorporate heterogeneity in the capture probabilities, we generate $P_{h}^{(l)}$ from two different choices of probability distributions S1: $Beta (2,2)$, and S2: $Beta (2,4)$ for $l=1,2,3$. We also generate data from Rasch model \citep{Rasch61} given by (B.3.1) in Section B.3 of the Supplementary Material, where the capture status of the \textit{h}-th individual in List $l$, $Z_{h}^{(l)}$ is a Bernoulli variate with probability $P_{h}^{(l)}=\frac{\exp[v_{h}+s_{l}]}{1+\exp[v_{h}+s_{l}]}$, for $h=1,\ldots,N$, $l=1,2,3$, with $v_{h}$'s as standard normal variates and two different choices of the fixed effects S3: $(s_{1},s_{2},s_{3})=(-1,0,1)$, and S4: $(s_{1},s_{2},s_{3})=(1,0.5,0.1)$. Note that all the aforementioned configurations (S1, S2, S3, S4) for data generation completely deviates from the assumed structure of the fitted models. The resulting RAMEs from the proposed THBM and its competitors are plotted in Figure \ref{RAME_S}. Following similar mechanism as in Section \ref{Simulation}, we present CP and LCI of the proposed and its competitors in Figure \ref{CP_S}. Here also, the proposed estimator outperforms the existing competitors with respect to RAME as discussed in Section \ref{Simulation}. Overall, the CPs of the proposed estimator and $M_{tb}$ model are close to $100\%$ and much better than all other estimators. In particular, the performance of SC is very poor with respect to RAMEs and CPs. As expected, the performance of the model IM is the worst in all respect.

\section{Applications}\label{Casestudy}
We apply our model to the two different case studies as mentioned in Section \ref{motivating}, one is a specific neurological disease among 1991 Gulf War veterans, and the other is on the disease surveillance among the survivors of 9/11 WTC Twin Towers Terrorist attack in USA. Both the datasets are presented in Table A.2 in the Supplementary Material. 	

\subsection{Analysis of Amyotrophic Lateral Sclerosis Count Data due to Gulf War 1991}\label{ALS}
First, we consider the case of Amyotrophic Lateral Sclerosis (ALS) disease among the 1991 Gulf War veteran military personnel. As per the report of \citet[Chapter 1]{Committee16-10}, nearly 697000 US troops were developed over the course of buildup and the war which started in August 1990 and ended by February, 1991. Besides injuries and deaths among the coalition forces, a large number of veterans suffered from various health-related issues during and after the war, which persisted for more than 25 years. From studies conducted in 1995 and later on, it has been found that the deployed Gulf War veterans suffered more than their non-deployed counterpart, both in numbers and severity \citep[Chapter 4]{Committee16-10}. In particular, $9.4\%$ of deployed veterans suffered from a gross neurological disorder, whereas the figure is $6.3\%$ for non-deployed personnel. ALS is a specific nervous system rare disease that weakens muscles and impacts physical function. In an epidemiological investigation regarding ALS among 1991 Gulf War veterans, it was a concern regarding possible under-ascertainment of the cases, especially among the non-deployed veterans who were hypothesized to receive less incentive to participate in the study \citep{Horner03}. In previous studies, some evidences of positive dependence among the sources are found especially for the non-deployed personnel. As pointed out by \citet{Gulf}, under-ascertainment could have a substantive effect on the rates and associated risk since ALS is a rare disease. Therefore, in this report, our prime interest is to estimate the size or under-ascertainment of the deployed and non-deployed veterans affected by ALS. In this context, another important issue was the diverse demographic profile of the military personnel, which was quite different in case of Gulf War compared to previous wars. In particular, the US military personnel in Gulf War were more ethnically diverse than their counterparts deployed in previous wars \citep[Chapter 1]{Committee16-10}. This phenomenon potentially induces a non-negligible amount of heterogeneity in the inclusion probabilities of veterans in different lists available for reporting ALS disease. In the original survey, four data sources of ALS cases based on capture-recapture experiment are (\textit{i}) Veterans Affairs database (V), (\textit{ii}) Department of Defense database (D), (\textit{iii}) Phone-line database (P), and (\textit{iv}) ALS Association database (A). Details of the study design and data sources can be found in \citet{Horner03,Gulf}. In total, 107  verified ALS cases were reported in the troops during the 10-year period beginning from August 2, 1990, and 40 out of these cases were from 696,118 deployed, and remaining 67 were from 1,786,215 non-deployed military personnel.
In the present study, we merge the data from last two sources P, A and rename it as `PA' to avoid the sparsity in the data from both the P and A sources. Hence, the TRS datasets, based on the three lists V, D and PA, for both the deployed and non-deployed veterans are presented in Table A.2 of the Supplementary Material.

In order to implement the proposed estimation methodology discussed in Section \ref{Inference}, we generate $K=1000$ samples from both the conditional distributions of $\boldsymbol{\mathcal{P}}$ and $\boldsymbol{y}$ to execute the Step $2$ (in Section \ref{MCEM}) and then, apply the EM algorithm using Steps 3 and 4 until convergence. We compute the estimate of the prevalence of ALS-affected deployed and non-deployed veterans and corresponding 95\% confidence intervals based on 1000 nonparametric bootstrap samples as it is done in simulation (section \ref{Simulation}). The same statistics are provided for the existing methods as considered in Section \ref{Simulation} to compare with the proposed one. The results are reported in Table \ref{Gulf_WTC_analysis}. The estimate of the ALS-affected veterans, based on THBM, is greater than that of all other methods under consideration. All the competitors, except model $M_{tb}$, yield similar estimates of the number of ALS cases for both the deployed and non-deployed veterans, and indicate a relatively low (moderate) rate of undercount\footnote{Rate of undercount is defined as $\frac{\hat{N}-x_{0}}{\hat{N}}$.} for deployed (non-deployed) personnel. 
Surprisingly, no undercount of ALS cases is observed for the deployed veterans based on model $M_{tb}$ in contrast to the findings of all other methods. The $95\%$ confidence intervals based on PQSM and LLM are extremely wide for both the deployed and non-deployed cases. \citet{Lum13} argued that large levels of uncertainty may imply drastic violations of assumptions and hence produce such unrealistic findings. The estimated undercount of ALS cases for the deployed veterans ($33\%$) is significantly higher than that of the non-deployed ($16\%$) personnel based on the proposed model. In total, 131 ALS cases are estimated based on THBM. We also compute the degree of list-dependence over the three sources V, D and PA based on the proposed model. The results indicates that the three sources are dependent among each other for $50\%$ ($41\%$) of the deployed (non-deployed) veterans. 
The coverage of ALS cases for both the deployed and non-deployed veterans in the V list is similar (56\% for deployed and 60\% for non-deployed cases). Like V lists, non-deployed cases are enlisted at a higher rate (62\%) in the D list compared to their deployed counterparts (34\%). However, ALS cases among the deployed (59\%) are more prevalent than that among the non-deployed personnel (42\%) in the combined list PA.
The average annual cumulative incidence rate (AACIR)\footnote{AACIR=$\frac{\text{Number of cases of a disease}}{\text{Number of years} \times \text{Number of persons at risk}}\times 100000$} of ALS among deployed and non-deployed military personnel were computed as 0.57 and 0.38 per 100,000 persons, respectively \citep{Horner03}. 
However, these figures did not account for the underreported cases, and any conclusion drawn from these statistics may be misleading. In particular, the relative risk of post-war ALS is underestimated when underreporting is higher among the Gulf war deployed compared to that of the non-deployed personnel. Therefore, it is essential to estimate the underreported cases and adjust the calculation of the AACI for an unbiased evaluation of the post-war syndromes. As per our estimate, we found that the AACI of ALS among Gulf war veterans increases significantly to 0.76 per 100,000 persons, and the same for non-deployed personnel increases marginally to 0.44 per 100,000 persons.	


\subsection{Analysis of World Trade Centre Occupants Data before the Terrorist Attack in 2001}

On September 11, 2001, the world was shocked by the news of an unprecedented terrorist attack on the Twin Towers of the World Trade Center, New York City. It was reported that 2,753 individuals were killed in and around the twin towers. Initially, a total of 2,750 deaths (excluding the terrorists) were confirmed at the World Trade Center site. Later, the New York City medical examiner's office enlisted those who died of illnesses caused by exposure to dust from the WTC site. 
Several reports gradually emerged after the incident by the organizations involved in the Public Health Care Response, which established that the deadliest attack had a direct physical and mental impact on the survivors.
For instance, 33,000 victims lost their lives after suffering from respiratory, gastrointestinal and mental health problems \citep{McAuliff2015}, and 1140 due to cancer from the exposure to toxins at Ground Zero \citep{Evans13}. Unfortunately, no official records have been kept on the casualties of the whole 9/11 attacks either by the FBI or the New York City administration in their respective crime records \citep{Hanrahan11}. As a result, proper data collection on the immediate death toll as well as subsequent surveillance on the health condition of the actual victims has not been done systematically over time. Thus, it is an absolute necessity to gain an idea about the actual count of the survivors to make effective policy decisions for their well-being of physical and mental health \citep{Klitzman03}. To understand the severity of such an act, computation of survival rate requires the knowledge on actual total count who were under the risk of death or injury due to that terrorist activity. 
In this article, we primarily confined ourselves for estimating the size of the survivors of the 9/11 WTC attack.
The World Trade Center Health Registry (WTCHR) database is prepared to compare the health condition of the people directly exposed to the WTC disaster with that of the general population \citep{WTCHR}. It helps to determine the extent of persistent physical and mental health issues and identify if any new symptoms and conditions have emerged. It comprises three sources - (\textit{i}) Self-identified individuals list (SI), (\textit{ii}) Businesses list (BL) and  (\textit{iii}) Port Authority list from New York and New Jersey Ports (PA). See \citet{Murphy09} for detail description of this dataset. \citet{Murphy09} pointed out the existence of \textit{list dependence} as well \textit{individual heterogeneity} in the available triple system data. 

Here, we consider the same estimation mechanism for all the models as considered in the case of the Gulf War case study in Section \ref{ALS}, and the results are presented in Table \ref{Gulf_WTC_analysis}. As per the proposed method, 13919 individuals were present in the twin towers when the first plane hit the north tower. This estimate is higher than all the existing competitors except the estimate based on PQSM. Our model indicates that the $35.6\%$ of survivors exhibits behavioural dependence among the three secures. To understand the severity of 9/11 WTC attack, computation of survival rate requires the knowledge on actual total count who were under the risk of death or injury due to that terrorist activity. Results based on THBM imply approximately $82.5\%$ of individuals survived this horrible incident. According to WTCHR, all these survivors were exposed to the risk of short-term and/or long-term physical and mental health effects \citep{WTCHR}.

\begin{table}[h]
	\tiny
	\begin{center}
		\caption{Summary results of data analysis on the ALS disease surveillance during 1991 Gulf War and WTC Terrorists attacks in 2001.}
		\label{Gulf_WTC_analysis}
		\begin{threeparttable}
			\resizebox{1\textwidth}{!}{
				\begin{tabular}{|lcccccc|}
					\hline
					&  THBM & SC & QSM & PQSM & LLM & $M_{tb}$\\
					\hline
					& \multicolumn{6}{c|}{Gulf War Deployed Veterans}\\
					$\hat{N}$ & 53 & 44 & 43 & 45 & 45 & 40\\
					CI & (42, 86) & (36, 52) &  (38, 79) & - & (36, 70) & (40, 88)\\
					& \multicolumn{6}{c|}{}\\
					& \multicolumn{6}{c|}{Gulf War Non-deployed Veterans}\\
					$\hat{N}$ & 78 & 74 & 70 & 72 & 72 & 134\\
					CI & (68, 107) & (69, 98) &  (67, 93) & (66, 162) & - & (68, 478)\\
					& \multicolumn{6}{c|}{}\\
					& \multicolumn{6}{c|}{WTC Terrorists attack}\\
					$\hat{N}$ & 13919 & 11977 & 11906 & 14698 & 12124 &  8974\\
					CI & (13733, 14112) & (11203, 12718) &  (11266, 12571) & (13015, 16092) & (10812, 12397) & (8973, 9197)
					\\
					\hline
				\end{tabular}
			}
			\begin{tablenotes}
				\tiny
				\item
				THBM: Trivariate Heterogeneous Bernoulli Model; SC: Sample Coverage;
				QSM: Quasi-symmetry Model;\\
				PQSM: Partial Quasi-symmetry Model;
				LLM: Log-linear Model, $M_{tb}$: Time Behavioural Response Variation Model 
				\
			\end{tablenotes}
		\end{threeparttable}
	\end{center}
\end{table}

\section{Discussions}\label{Discussion}
In this paper, the issue of population size estimation motivated by real case studies on the 9/11 terrorists attack on WTC, USA and the Gulf war survivors has been addressed. For this purpose, a novel model for capture-recapture data is proposed and an MCEM estimation methodology has been developed. The proposed method seems to have an edge in terms of flexibility in modeling both the \textit{list-dependence} and heterogeneous catchability. For example, the proposed method provides a clear picture of the nature and the extent of behavioral dependence of the individuals in the population under consideration. Moreover, the estimation methodology avoids implausible restrictions on parameters and automatically handles the identifiability issues with respect to TRS data structure. The simulation study exhibits robustness of the proposed method under model misspecification and superiority of performance over the existing models.

In the context of disease monitoring, we often emphasize the necessity to correct the conventional method of calculating incidence rate adjusting the estimated size of the under-reported cases for an unbiased evaluation of post-war syndromes. Based on our analyses, we found that the average annual cumulative incidence of ALS increases significantly among the Gulf war veterans after adjusting the undercount events. In the context of the WTC terrorist attack, several reports indicated that the survivors, including rescuers, police personnel and other government officials, are more or less affected by mental and environmental health issues \citep{Klitzman03, McAuliff2015}. Some of them died after the incident, and the rest carried their losses in their body or mind forever. Our estimate indicates higher undercount rate compared to the previous studies. However, there is not much work available on the risk estimate of post-attack diseases with suitable adjustments for the undercount. The lack of demographic information for this case study hampered our ability to carry out further analysis to identify the risk factors. 

The scope of the proposed model goes far beyond the particular case studies under consideration and have much wider applicability in the fields of public health, demography, social sciences, etc. Recently, multiple systems estimation strategies have been applied to estimate the number of victims of human trafficking and modern slavery. In such applications, sparse or even no overlapping lists poses challenges in model fitting \citep{Chan2021}. The proposed approach can be extended to model such datasets to address the existing estimation issues.

\section*{\small Acknowledgement}
The authors also like to thank Mr. Debjit Majumder for his help in creating the bar diagrams. The work of Dr. Prajamitra Bhuyan is supported in part by the Lloyd’s Register Foundation programme on data-centric engineering at the Alan Turing Institute, UK. The work of Dr. Kiranmoy Chatterjee is supported by the Core Research Grant (CRG/2019/003204) on Mathematical Sciences by the Science and Engineering Research Board, Department of Science \& Technology, Government of India.

\section*{\small Supplementary Information}
Supplementary material is openly available at \doi{10.13140/RG.2.2.22118.19525}.

\bibliographystyle{apalike}
\bibliography{Bibliography_TBM}

\end{document}


\appendix
	
	\begin{center}
		\Large \bf Supplementary Material for ``Estimating Prevalence of Post-war Health Disorders Using Capture-recapture Data'' by Bhuyan and Chatterjee
	\end{center}

	\setcounter{page}{1}
	
	\section{Real Datasets of the Motivating Case Studies in TRS Data Structure }\label{Dataset}
	\renewcommand{\thetable}{\thesection.\arabic{table}}

	\renewcommand\thefigure{\thesection.\arabic{figure}} 
	
	\begin{table}[ht]
		\tiny
		\centering
		\caption{TRS Datasets on ALS disease among the Gulf war veterans and WTC Twin Tower population at the time of 9/11 terrorist attack.}
		\resizebox{5.6in}{!}{
			\begin{tabular}{|cccccccccc|}
				\hline
				Dataset & Stratum &	$x_{111}$ & $x_{110}$ & $x_{101}$ & $x_{011}$ & $x_{100}$ & $x_{010}$ & $x_{001}$ & Total ($x_0$)\\
				\hline
				ALS & All &	24 & 23 & 19 & 9 & 10 & 10 & 12 & 107 \\
				& Deployed & 10 & 2 & 12 & 4 & 5 & 2 & 5 & 40\\
				& Non-deployed & 14 & 21 & 7 & 5 & 5 & 8 & 7 & 67 \\
				\multicolumn{10}{|c|}{}\\
				WTC &  &	174 & 88 & 1658 & 750 & 1702 & 270 & 4323 & 8965 \\
				\hline
			\end{tabular}
		}
		\label{case_Data}
	\end{table}
	
	\section{Existing Models and Estimates}\label{existing}
	\numberwithin{equation}{subsection}
	\setcounter{table}{0}
	\renewcommand{\thetable}{\thesection.\arabic{table}}
	
	\setcounter{figure}{0}
	\renewcommand\thefigure{\thesection.\arabic{figure}} 
	
	In this section, we present frequently used models for TRS and associated estimates for the population size. These models incorporate \textit{individual heterogeneity} and/or \textit{list-dependence}. The assumptions associated  with these models and their limitations in real applications are also discussed briefly. We consider these estimates to compare the performance of the proposed method through simulation study discussed in Section 4 of the main article.
	
	\subsection{Log-linear Model}
	\citet{Fienberg72b} and \citet{Bishop75} discussed several log-linear models (LLMs) to account for \textit{list-dependence} using interaction effects. The general LLM under TRS is  given by
	\begin{eqnarray}
		\log(m_{ijk})=u_0 + u_{1(i)} + u_{2(j)} + u_{3(k)} + u_{12(ij)} + u_{13(ik)} + u_{23(ik)} + u_{123(ijk)},\label{log-linear_gen}
	\end{eqnarray} 
	where $m_{ijk}=\mathbb{E}[x_{ijk}]$, $u_{l(0)}+u_{l(1)}=0$,  $u_{l l^{\prime} (0j)}+u_{l l^{\prime}(1j)}=0$,
	$u_{l l^{\prime} (i0)}+u_{l l^{\prime}(i1)}=0$,
	$u_{l l^{\prime}l^{\ast} (0jk)}+u_{l l^{\prime}l^{\ast}(1jk)}=0$,
	$u_{l l^{\prime}l^{\ast} (i0k)}+u_{l l^{\prime}l^{\ast}(i1k)}=0$,
	$u_{l l^{\prime}l^{\ast} (ij0)}+u_{l l^{\prime}l^{\ast}(ij1)}=0$,
	for all $l, l^{\prime},l^{\ast}=1,2,3$, with $l\ne l^{\prime}\ne l^{\ast}$ \citep[p.64]{Bishop75}. The parameters $u_l$, $u_{l l^{\prime}}$ and $u_{123}$ denote the main effects, pairwise interaction effects and the second order interaction effect for $l,l^{\prime}=1,2,3$. See \citet{Fienberg72a} and \citet{Fienberg72b} for more details. To ensure estimability of the model given in (\ref{log-linear_gen}), one needs to consider $u_{123}=0$, i.e. no second order interaction between lists. Under this assumption, the estimate of $m_{000}$ is given by
	\begin{eqnarray}
		\hat{m}_{000}=\frac{\hat{m}_{111}\hat{m}_{001}\hat{m}_{100}\hat{m}_{010}}{\hat{m}_{101}\hat{m}_{011}\hat{m}_{110}},\nonumber
		\label{est_no_2nd_order}
	\end{eqnarray}
	where $\hat{m}_{ijk}$ is the maximum likelihood estimate of $m_{ijk}$ assuming $x_{ijk}$ to be a realization of an independent Poisson random variate for all $(i,j,k)$th cells, except the $(0,0,0)$th cell. Finally, the estimate of $N$ is obtained as $\hat{N}_{LLM}=x_0+\hat{m}_{000}$ \citep{Fienberg72b}. This is equivalent to the estimator proposed by \citet{Zaslavsky93} with $\alpha_{EPA}=1$. With additional assumption that $u_{12}=u_{13}=u_{23}=0$, LLM in (\ref{log-linear_gen}) reduces to IM.

	\subsection{Time Behavioural Response Variation Model: $M_{tb}$}\label{Mtb}
	It is often observed in TRS that an individual's behavior changes in subsequent recapture attempts after the initial attempt. This change is known as behavioral response variation. When this behavioral response variation is considered along with the assumption of list variation in capture probabilities, one would have a model known as $M_{tb}$ model \citep{Otis78, Chatterjee16c}. However, $M_{tb}$ model does not account for the \textit{individual heterogeneity} in capture probabilities.
	
	Denote the first-time capture probability of any individual in the $l$-th list is by $f_l$ for $l=1, 2, 3$, and the recapture probability is denoted by $c_l$ for $l=2, 3$. Further, $u_l$ and $m_l$ denote, respectively, the number of first-time captured and recaptured individuals in the $l$-th list. Therefore, based on the the sufficient statistics ($u_1=x_{1\cdot\cdot},u_2=x_{01\cdot},u_3=x_{001},m_2=x_{11\cdot}, m_3=x_{101}+x_{011}+x_{111}$) and the assumption of constant proportionality, i.e., $c_l/f_l=\phi,$ for $l=2,3$, likelihood of the model $M_{tb}$ is given by
	\begin{eqnarray}
		L(N,\textbf{\textit{f}},\phi)=\frac{N!}{(N-x_0)}f_1^{u_1}(1-f_1)^{N-u_1}\phi^{m_2+m_3}\prod_{l=2}^{3}f_l^{u_l+m_l}(1-f_l)^{N-M_{l+1}}(1-\phi p_l)^{M_l-m_l},\nonumber
	\end{eqnarray}
	where $x_0=\sum_{i,j,k: ijk\neq000}x_{ijk}$ and $M_l=u_1+u_2+...+u_{l-1}$ denote the total number of distinct captured individuals and the number of individuals captured at least once prior to the $l$-th attempt, respectively \citep{Chatterjee20b}. In this context, \citet{Chao00} discussed various likelihood based methods for estimation of $N$. Note that, $M_{tb}$ model does not utilize full information available from seven known cells in TRS, and hence, it looses efficiency in order to estimate $N$ \citep{Chatterjee20b}. Moreover, the assumption of $c_2/f_2=c_3/f_3=\phi$ is not justifiable in various real life applications.

	\subsection{Models with individual heterogeneity}\label{Hetero} 
	In educational statistics and Psychometry, Rasch model \citep{Rasch61} is widely is used to account for the \textit{individual heterogeneity} and list variation. The capture probabilities are model using logistic regression:
	
	\begin{eqnarray}
		\log\left[\frac{\mathbb{E}[Z_h^{(l)}]}{1-\mathbb{E}[Z_h^{(l)}]}\right]=v_h+s_l, \hspace{0.3in} h=1,2,\cdots,N;\hspace{0.2in}l=1,2,3,\label{Rasch}
	\end{eqnarray} 
	where $v_h$ is a random effect representing the catchability of $h$th individual and $s_l$ is effect of $l$th list. Even with the absence of behavioural response variation, dependence among the lists are induced is purely due to \textit{individual heterogeneity} \citep{Chao01a}.  When we set $v_h=0$ in (\ref{Rasch}), the model reduces to multiple-recapture model with list independence. We now represent the Rasch model as a mixed effects generalized linear model that allows for \textit{individual heterogeneity} and list variation in the following subsections.

	\subsubsection{Quasi-symmetry Model}
	The marginal cell probabilities for all the eight cells in a TRS can be represented as:
	\begin{eqnarray}
		\log(p_{ijk})=a+is_1+js_2+ks_3+\gamma(i_+),\label{quasi}
	\end{eqnarray} 
	where $i_+=i+j+k$ and $\gamma(t)=\log\left[\mathbb{E}\left[\exp (v_h t)|i=j=k=0\right]\right]$. See \citet{Darroch93, Fienberg99} for detailed derivation. Note that the term $\gamma(i_+)$ is invariant with respect to permutations of ($i,j,k$), and hence the model in (\ref{quasi}) is known as quasi-symmetry model (QSM) \citep{Darroch81}. This is equivalent to the following assumption involving two constraints
	\begin{eqnarray}
		p_{011}p_{100}=p_{101}p_{010}=p_{110}p_{001}\label{two-constraints}
	\end{eqnarray}
	which do not involve $p_{000}$. Therefore, an additional assumption is required, such as no second-order interaction exists. Under this assumption, the estimated count of the $(0,0,0)$th cell is given by
	\begin{eqnarray}
		\hat{x}_{000}=\frac{\hat{x}_{111}\hat{x}_{100}\hat{x}_{010}\hat{x}_{001}}{\hat{x}_{110}\hat{x}_{101}\hat{x}_{011}},\nonumber\label{no-2nd-order}
	\end{eqnarray}
	where $\hat{x}_{ijk}$ is the maximum likelihood estimate of $\mathbb{E}[x_{ijk}|x_0]$ \citep{Darroch93}. 
	
	\subsubsection{Partial Quasi-symmetry Model}
	In many situations, the constraint provided in (\ref{two-constraints}) associated with quasi-symmetry model is not realistic.  However, it is reasonable to assume 
	$p_{011}p_{100}=p_{101}p_{010}$
	if $x_{011}x_{100}$ and $x_{101}x_{010}$ are close enough \citep{Darroch93}. This condition implies that the patterns of \textit{individual heterogeneity} are different for two different group of sources or lists as considered in generalized Rasch model proposed by \citet{Stegelman83}. Under this assumption,
	the marginal cell probabilities can be represented as:
	\begin{eqnarray}\label{partial-quasi}
		\log(p_{ijk})=a+is_1+js_2+ks_3+\gamma(i+j,k), \nonumber
	\end{eqnarray} 
	where $\gamma(t_1,t_2)=\log\left[\mathbb{E}\left[\exp(v_ht_1)\exp(z_h t_2)|i=j=k=0\right]\right]$ with two nonidentiacally distributed random effects $v_h$ and $z_h$ associated with group of lists ($L_1$, $L_2$) and $L_3$, respectively . This model in is known as partial quasi-symmetry model (PQSM). See \citet{Darroch93, Fienberg99} for more details.
	
	\subsection{Sample Coverage Method}\label{SCM}
	The idea of the sample coverage of a given sample is the probability weighted fraction of the population captured in that sample. The main principle relies on the fact that it is difficult to estimate the number of undercount directly, but the sample coverage can be well estimated in any non-independent situation with heterogeneous capture probability \citep{Chao98}. In the context of TRS, the sample coverage is defined based on three hypothetical populations I, II, and III with capture probabilities $\left\{\mathbb{E}\left[Z_h^{(1)}\mid Z_h^{(2)},Z_h^{(3)}\right]; h=1,\ldots, N \right\}$, $\left\{\mathbb{E}\left[Z_h^{(2)} \mid Z_h^{(1)},Z_h^{(3)}\right]; h=1,\ldots, N \right\}$, and \\
	$\left\{\mathbb{E}\left[Z_h^{(3)} \mid Z_h^{(1)},Z_h^{(2)}\right]; h=1,\ldots, N \right\}$ corresponding to lists $L_1$, $L_2$, and $L_3$, respectively. In particular, the sample coverage of first two lists $L_1\cup L_2$ with respect to the population III is given by
	\begin{eqnarray}
		C_{III}(L_1\cup L_2)=\frac{\sum_{h=1}^{N}\mathbb{E}\left[Z_h^{(3)} \mid Z_h^{(1)},Z_h^{(2)}\right]\mathbb{I}[Z_h^{(1)}+Z_h^{(2)}>0]}{\sum_{h=1}^{N}\mathbb{E}\left[Z_h^{(3)} \mid Z_h^{(1)},Z_h^{(2)}\right]}. \nonumber
	\end{eqnarray}
	Similarly, one can obtain the sample coverages of other two possibles combinations of samples, $C_{II}(L_1\cup L_3)$ and $C_{I}(L_2\cup L_3)$ with respect to the populations II and I, respectively. Now, the sample coverage of the three lists is defined as $C=\frac{1}{3}\left\{C_{III}(L_1\cup L_2)+C_{II}(L_1\cup L_3)+C_{I}(L_2\cup L_3)\right\}$, and estimated by $\hat{C}=1-\frac{1}{3}\left[\frac{x_{100}}{n_1}+\frac{x_{010}}{n_2}+\frac{x_{001}}{n_3}\right]$. Finally, the estimator of $N$ on based sample coverage approach is given by
	\begin{eqnarray}
		\hat{N}_{sc}=\frac{x_{\cdot 11}+x_{1\cdot 1}+x_{11\cdot}}{3\hat{C}}\left\{1-\frac{1}{3\hat{C}}\left[
		\frac{(x_{1\cdot 0}+x_{\cdot 10})x_{11\cdot}}{n_1n_2}+
		\frac{(x_{10\cdot}+x_{\cdot 01})x_{1\cdot 1}}{n_1n_3}+
		\frac{(x_{0\cdot 1}+x_{01\cdot})x_{\cdot 11}}{n_2n_3}\right]\right\}^{-1}.\nonumber
	\end{eqnarray}
	We refer this estimate as SC. From empirical studies, it is found that the performance of $\hat{N}_{sc}$ is satisfactory if the sample coverage is over 55\%. The performance also depends on the remainder term involved in the derivation of $\hat{N}_{sc}$, and the bias increases with its magnitude \citep{Chao98}. Moreover, this method may produce infeasible estimates, i.e. $\hat{N}_{sc}< x_{0}$, unlike other methods.

	\numberwithin{equation}{subsection}
	\setcounter{table}{0}
	\renewcommand{\thetable}{\thesection.\arabic{table}}
	
	\setcounter{figure}{0}
	\renewcommand\thefigure{\thesection.\arabic{figure}} 
	
	\section{Derivations of Distributions of the Latent Cell Counts}\label{Derev_y}
	The conditional density of the latent cell counts $\boldsymbol{y}$, given $(N,\boldsymbol{\alpha},\boldsymbol{\mathcal{P}},\boldsymbol{x})$, is easily obtained from (4) as
	$$
	\pi(\boldsymbol{y}|N,\boldsymbol{\alpha},\boldsymbol{\mathcal{P}},\boldsymbol{x})\propto \prod_{i,j,k=0,1}\pi(y_{ijk}|\boldsymbol{\theta},\boldsymbol{\mathcal{P}},\boldsymbol{x}),
	$$
	where
	\begin{eqnarray}
		\pi(\boldsymbol{y_{111}}|\boldsymbol{\theta},\boldsymbol{x}) & \propto & Multinomial(x_{111};Q_{111,1}^*,Q_{111,2}^*,Q_{111,3}^*,Q_{111,4}^*,Q_{111,5}^*),\nonumber\\
		\pi(y_{110,1}|\boldsymbol{\theta},\boldsymbol{x}) & \propto & Binomial(x_{110};(1-\alpha_0) \mathcal{P}_1 \mathcal{P}_2 (1-\mathcal{P}_3)),\nonumber\\
		\pi(y_{011,1}|\boldsymbol{\theta},\boldsymbol{x}) & \propto & Binomial(x_{011};(1-\alpha_0)(1-\mathcal{P}_1)\mathcal{P}_2 \mathcal{P}_3),\nonumber\\
		\pi(y_{100,1}|\boldsymbol{\theta},\boldsymbol{x}) & \propto & Binomial(x_{100};(1-\alpha_0)\mathcal{P}_1(1-\mathcal{P}_2)(1-\mathcal{P}_3)),\nonumber\\
		\pi(y_{101,1}|\boldsymbol{\theta},\boldsymbol{x}) & \propto & Binomial(x_{101};(1-\alpha_0)\mathcal{P}_1(1-\mathcal{P}_2)\mathcal{P}_3),\nonumber\\
		\pi(y_{010,1}|\boldsymbol{\theta},\boldsymbol{x}) & \propto & Binomial(x_{010};(1-\alpha_0)(1-\mathcal{P}_1)\mathcal{P}_2(1-\mathcal{P}_3)),\nonumber\\
		\pi(y_{001,1}|\boldsymbol{\theta},\boldsymbol{x}) & \propto & Binomial\left(x_{001};(1-\alpha_0)(1-\mathcal{P}_1)(1-\mathcal{P}_2)\mathcal{P}_3\right),\nonumber\\
		\pi(\boldsymbol{y_{000}}|\boldsymbol{\theta},\boldsymbol{x}) & \propto & Multinomial(N-x_{0};Q_{000,1}^*,Q_{000,2}^*,Q_{000,3}^*,Q_{000,4}^*,Q_{000,5}^*)\nonumber,\label{sampling_y}
	\end{eqnarray}   
	with
	$\boldsymbol{y_{111}}=\left(y_{111,1},y_{111,2},y_{111,3},y_{111,4},y_{111,5}\right)$, $\boldsymbol{y_{000}}=\left(y_{000,1},y_{000,2},y_{000,3},y_{000,4},y_{000,5}\right)$,
	$Q_{111,1}^*=(1-\alpha_0)\mathcal{P}_1 \mathcal{P}_2 \mathcal{P}_3/Q_{111}$, $Q_{111,2}^*=\alpha_1 \mathcal{P}_1 \mathcal{P}_3/Q_{111} $, $Q_{111,3}^*=\alpha_2 \mathcal{P}_1 \mathcal{P}_2/Q_{111}$, $Q_{111,4}^*=\alpha_3 \mathcal{P}_1 \mathcal{P}_2/Q_{111}$, $Q_{111,5}^*=\alpha_4\mathcal{P}_1/Q_{111},$ $Q_{111}=(1-\alpha_0)\mathcal{P}_1 \mathcal{P}_2 \mathcal{P}_3+\alpha_1 \mathcal{P}_1 \mathcal{P}_3+\alpha_2 \mathcal{P}_1 \mathcal{P}_2+\alpha_3 \mathcal{P}_1 \mathcal{P}_2+\alpha_4\mathcal{P}_1$;\\
	$Q_{000,1}^*=(1-\alpha_0) (1-\mathcal{P}_1) (1-\mathcal{P}_2) (1-\mathcal{P}_3)/Q_{000}$, $Q_{000,2}^*=\alpha_1 (1-\mathcal{P}_1) (1-\mathcal{P}_3)/Q_{000}$, $Q_{000,3}^*=\alpha_2 (1-\mathcal{P}_1) (1-\mathcal{P}_2)/Q_{000}$, $Q_{000,4}^*=\alpha_3 (1-\mathcal{P}_1) (1-\mathcal{P}_2)/Q_{000}, Q_{000,5}^*=\alpha_4(1-\mathcal{P}_1)/Q_{000},$ $Q_{000}=
	{(1-\alpha_0) (1-\mathcal{P}_1) (1-\mathcal{P}_2) (1-\mathcal{P}_3)}+\alpha_1 (1-\mathcal{P}_1) (1-\mathcal{P}_3)+\alpha_2 (1-\mathcal{P}_1) (1-\mathcal{P}_2)+{\alpha_3 (1-\mathcal{P}_1) (1-\mathcal{P}_2)}+\alpha_4(1-\mathcal{P}_1)$. 
	

	\bibliographystyle{apalike}
	\bibliography{Bibliography_TBM}